\documentclass[12pt,epsf]{article}
\usepackage{graphicx, amsmath}

\begin{document}

\sloppy
\begin{flushright}{SIT-HEP/TM-31}
\end{flushright}
\vskip 1.5 truecm
\centerline{\large{\bf  Vacuum energy reduction through destabilization:}}
\centerline{\large{\bf a unification of quintessence and a dynamical
approach?}}   
\vskip .75 truecm
\centerline{\bf Tomohiro Matsuda
\footnote{matsuda@sit.ac.jp}}
\vskip .4 truecm
\centerline {\it Laboratory of Physics, Saitama Institute of
 Technology,}
\centerline {\it Fusaiji, Okabe-machi, Saitama 369-0293, 
Japan}
\vskip 1. truecm
\makeatletter
\@addtoreset{equation}{section}
\def\theequation{\thesection.\arabic{equation}}
\makeatother

\begin{abstract}
\hspace*{\parindent}
Although quintessence cosmologies seem to explain the amount of
cosmological constant today, the required conditions are severe.
For example, an extremely slowly varying and light scalar field that
rolls toward the vanishing vacuum energy state is needed, which
must satisfy the required initial conditions so that the energy density in
matter is comparable to the energy density stored in quintessence today.
Our first question is whether it is possible to explain why
quintessence ``knows'' that the Universe is going toward the vanishing
energy state.
We show from a new perspective that the conventional $O(H)$ corrections
 may reproduce the required potential.
Our second question is whether it is possible to obtain plausible amount
of the potential energy density today without tuning initial
conditions.
Since our resultant potentials mimic quintessence,
one may solve the coincidence problem using the conventional techniques 
for solving the coincidence problem in quintessence cosmology. 
However, in our scenario we have another option, which is qualitatively
 different from the one that have been 
 discussed in quintessence cosmology.  
Besides quintessence cosmology, we show that there is a relation between
 a dynamical approach, which suggests that we can solve the coincidence
 problem in a dynamical approach. 
As far as we know, this is the first attempt to discuss about
the unified picture and the important relations between these 
distinctive scenarios.
\end{abstract}

\newpage
\section{Introduction}
The existence of a cosmological constant that is driving the 
acceleration of the present Universe is perhaps the most striking
result that has been obtained in modern cosmological observations.
Related to the existence of a cosmological constant, there is a serious 
problem; we must explain the value of the energy density
stored in the cosmological constant today, which has to be comparable to
the critical density.
Although many possibilities have been discussed by many
authors\footnote{See ref.\cite{review-Quin} and references therein.},
it may still be fair to say that there is no successful
explanation as to why the cosmological constant appears at the present
size.
Among many proposals, quintessence has been proposed as the candidate of
the varying cosmological constant, which is a dynamical, slowly evolving
component with negative pressure.
A key problem with the quintessence proposal would be explaining why the
energy density stored in the quintessence and the matter energy densities
are comparable today.
There are two major aspects to the above coincidence problem; one is
that the initial condition in the early Universe has to be tuned so that
the energy 
densities become comparable today, and the other is that the
quintessence energy density is very tiny compared to typical scale of
particle physics.
The former condition may be softened by introducing a form of quintessence
called tracker fields, and the latter may be forced by direct
measurements.

Although quintessence cosmology seems attractive, it still needs further
explanations so as to why quintessence is an extremely slowly varying and
light scalar field  that rolls toward the vanishing energy state.
One may find a kind of solution in the string theory or brane models,
however it is still very difficult especially to explain why the
quintessence is 
rolling down the potential toward the vacuum with vanishingly small
energy density.
In this paper, we will consider the above questions trying to find 
explanations why the quintessence has its peculiar properties.
We show from a new perspective that the conventional $O(H)$ corrections
 may explain both (1) the smallness of the potential energy density
 and (2) the coincidence between the potential energy density and the
 energy density in matter.\footnote{Here we use the word ``$O(H)$
 corrections'' to mean the effective terms that appear with $H^n$, where
 $H$ is the Hubble constant.}
 Starting with simple setups, we will ``obtain''  potentials that mimic
quintessence. 

Moreover, we will show how one can generalize the conventional
quintessence cosmology so that it may accommodate itself to the
``coincidence'' in the present Universe. 

Our mechanism can be used to solve the coincidence problem in a
dynamical approach.

\section{Eta-problem and $O(H)$ corrections}
Let us briefly review $O(H)$ corrections that appear in scalar
potential.
For example, there are many theories where one has effective mass with 
$|\Delta m^2|=O(H^2)$.
If the field $\phi$ non-minimally couples to gravity, it may
acquire a correction to the mass squared $|\Delta m^2| \simeq \xi R$,
where $R$ is the curvature scalar.
For the popular choice of conformal coupling $\xi = 1/6$,
one obtains $|\Delta m^2| = 2 H^2$. 
A similar situation appears in cosmological models with supergravity,
where scalar fields acquire correction 
$|\Delta m^2| \simeq  H^2$.
A similar correction appears also in brane inflationary models, where
light scalar fields may acquire correction to the mass squared due to
the mechanism that stabilizes the internal space. 
Since these corrections are not favorable in obtaining successful
inflationary expansion in the early Universe, the appearance of the
above terms is called ``eta-problem''.  

Before we discuss about our mechanisms, we will explain some details
about the $O(H)$ corrections.
To be precise, the conventional ``$O(H)$ corrections'' do not always
mean the corrections of $O(H)$.
For example;
\begin{itemize}
\item In conventional supergravity, ``$O(H)$ corrections'' are 
      corrections from $F$-terms, not from the Hubble constant.
      Of course, they may coincide if the $F$-term dominates the Hubble
      constant, however we must be careful about this statement.
      This idea is used in $D$-term inflation models, since the ``$O(H)$
      corrections'' do not appear (or much smaller than the Hubble
      constant) if the dominant part of the inflationary potential is
      given by $D$-terms.
\item There may be corrections from the thermal plasma, which is
      given by
      \begin{equation}
       |\Delta m^2| \propto \frac{T^4}{M_p^2}.
      \end{equation}
      This term mimics ``$O(H)$ corrections'' if the thermal plasma
      dominates 
      the Universe. 
\item In KKLT models, flat directions may obtain $O(H)$ masses due to 
      the mechanism that stabilizes the extra dimensions.
\end{itemize}
Although the magnitude of these corrections may be of the same order,
these effects are qualitatively different from each other and may have
different coefficients. 

From the above arguments, it will be natural to introduce a new
definition of the so-called ``$O(H)$ corrections'', introducing
a new variable $H_s$, which is given by
\begin{equation}
H_s^2 \equiv \left(A_r \rho_r + A_m \rho_m + A_v \rho_v\right)/M_p^2,
\end{equation}
where $A_{r,m,v}$ are the model-dependent coefficients, as we have
mentioned above.
Here $\rho_r$, $\rho_m$ and $\rho_v$ are the energy density in
radiation, matter, and potential, respectively.
In this paper, we will consider ``$O(H_s)$ corrections'' instead of the
usual $O(H)$ corrections.
In our argument, the coefficients $A_{r,m,v}$ may take either sign, since
each correction may appear independently.\footnote{Be sure that we are
discussing about possibilities. It is still very hard to calculate the
coefficients.}

In addition to the effective mass terms that we have
discussed above, one may include higher corrections 
\begin{eqnarray}
\Delta m^2(H_s, \phi) &=& H_s^2 f_1\left(\frac{\phi}{M_*}\right)
+ H_s^4 f_2\left(\frac{\phi}{M_*}\right) + ...,
\end{eqnarray}
which may be given by
\begin{eqnarray}
\Delta m^2(H_s, \phi) &=& H_s^2 \left(B_{1,0} + B_{1,1}
\frac{\phi^{2}}{M_*^{2}} + ... + 
\frac{B_{1,i_*} \phi^{2i_*}}{M_*^{2i_*}}\right)\nonumber\\
&&+ H_s^4 \left(\frac{B_{2,0}}{M_*^2} + B_{2,1}
\frac{\phi^{2}}{M_*^{4}} + ... + 
\frac{B_{2,i_*} \phi^{2i_*}}{M_*^{2(i_*+1)}}\right)
 + ...,
\end{eqnarray}
where $B$'s are model-dependent coefficients, and 
$i_*$, $j_*$ are cut-off parameters.
We introduced $M_*$, which denotes the cut-off scale of the effective
theory.
One may take $M_*\simeq M_p$ for simplicity.

\section{Destabilized potential} 
Before we go into details of the mechanism, it will be better to 
discuss more about the situations that we will consider in this paper.
The important component in our mechanism is the scalar field $\phi$
whose potential is destabilized by $O(H)$ corrections.
As a result, the potential energy density is reduced when the field
$\phi$ rolls down the potential. 

First, let us consider a scalar field $\phi$, which is flat at the tree
level.
The potential for the field $\phi$ is given by the simplest form,
\begin{equation}
V(\phi)_{tree} = V_0.
\end{equation}
It is not surprising if the above potential is destabilized when 
the field $\phi$ obtains $O(H_s)$ corrections, which is given by
\begin{equation}
\label{correction1}
\Delta V(\phi) =  - \frac{1}{2}H_s^2 \phi^2,
\end{equation}
where $A_v > 0$ is assumed.
As we are considering $H_s$ instead of $H$, the correction does depend
on the coefficients $A_{r,m,v}$. 
If the the above correction appears with the negative sign, as we have
shown above in eq.(\ref{correction1}), the potential $V(\phi)$
is destabilized, and then, the field $\phi$ starts to roll down the
potential. 

Let us consider the evolution of the field $\phi$ in a region
$0< \phi < M_*$.
As the effective mass term  (\ref{correction1}) destabilizes the
potential, 
the field $\phi$ rolls away to reduce the total potential energy density.
In this region, the potential $V(\phi)$ is given by
\begin{eqnarray}
\label{int1}
V(\phi) &=& V_0 + \int_0^{\phi} 
\left[\frac{d }{d\phi}\Delta V\right] d\phi\nonumber\\  
&\simeq& V_0 - \frac{1}{2}A_v V(\phi)M_p^{-2} \phi^2,
\end{eqnarray}
where we have assumed that the vacuum energy dominates $H_s$;
\begin{equation}
H_s^2 \simeq A_v V(\phi)/M_p^2.
\end{equation}
From the above expression, we obtained the form of the potential,
which is given by
\begin{equation}
V(\phi) \simeq \frac{V_0}{1+\frac{A_v\phi^2}{2M_p^2}}.
\end{equation}
If the effective mass of the field $\phi$ is small,
the approximate form of the above potential for $A_v < 1$ and 
$\phi \ll M_p$ is given by
\begin{equation}
V(\phi) \simeq V_0\left(1 -\frac{A_v \phi^2}{2M_p^2}\right) \simeq V_0-
\frac{1}{2}H_s|_{\phi=0}\phi^2.
\end{equation}
On the other hand, if the potential is very steep (i.e. $A_v \gg 1$),
 the approximate form of the above potential in the region 
 $M_* >\phi > M_p\sqrt{2/A_v}$ is given by  
\begin{equation}
V(\phi) \simeq \frac{2M_p^2 V_0}{A_v}\phi^{-2},
\end{equation}
if the region exists.

As we will discuss later in this section, the above form of  $\Delta V$
may be altered in the outer region ($\phi > M_*$) where  
higher terms may start to dominate the potential.

As we are considering a scalar field $\phi$ that rolls down the
potential with the negative mass term (\ref{correction1}), the
roll of the field $\phi$ may satisfy the condition for fast-roll
inflation. 
To be more precise, if the initial value of $\phi$ is small and the
coefficient satisfies the condition $A_v \le\frac{3}{4}$, 
there will be a significant amount of fast-roll
inflation\cite{fast-roll-original, Dimopoulos-Fast-Osc} that lasts until
$\phi = M_*$, where higher terms may become significant.
We found that the total number of e-foldings corresponding to fast-roll
inflation is given by 
\begin{equation}
N = \frac{1}{F}\ln\left(\frac{M_*}{\phi_0}\right),
\end{equation}
where $\phi_0^2 \ge H^2 \sim V_0/M_p^2$, and the coefficient $F$ is
given by 
\begin{equation}
F=\frac{3}{2}\left[1-\sqrt{1-\frac{4}{3}|A_v|}\right] < \frac{3}{2}.
\end{equation}
Unlike the conventional scenario of fast-roll inflation,
the effective mass is not a constant in our model.
The effective mass decreases with time, since we are considering the
mass term that comes from the $O(H_s)$ correction.

If one is considering a theory with
many flat directions, the above destabilization may occur for the
large number of $n\gg 1$ scalar fields.
Then, assuming (for simplicity) spherical symmetry for $n$ scalar
fields, the factor of $n \gg 1$ appears in (\ref{int1}) as 
\begin{equation}
\label{int2}
V(\phi) = V_0 + n\int_0^{\phi} \left[\frac{d }{d\phi}\Delta V\right]
d\phi.
\end{equation}
We obtained the approximate form of the potential, which is
given by
\begin{equation}
V(\phi) \simeq \frac{V_0}{1+\frac{n A_v\phi^2}{2M_p^2}}.
\end{equation}
In this case, even if the potential satisfies the fast-roll condition
for each field, the reduction of the vacuum energy occurs much faster
than the single-field potential.

What is important here is the qualitative property that the field $\phi$
keeps on rolling as far as the effective mass term appears with the 
negative sign, while the driving force disappears when $H_s \simeq 0$ at
a distance.
Of course, the vacuum energy cannot vanish because the field becomes
extremely light before it crosses $H_s = 0$.
In our model, the peculiar field that ``knows'' the zero-energy vacuum
is not required, but it is obtained from the conventional effective
potential.
This property is distinguishable.
In fact, any field can accommodate itself to the vanishing potential
energy if only it satisfies the above conditions.

Let us discuss more about the destabilized potential $V(\phi)$.
The above correction (\ref{correction1}) is plausible within the region
$0< \phi < M_*$, where $M_*$ is a cut-off scale.
Above the cut-off scale $M_*$, higher corrections will be
important.
Then, for example in a region $M_* < \phi < \phi_{1}$, where the value
of $\phi_1$ is model-dependent, we can assume that the
potential is dominated by a term coming from the higher
corrections
\begin{equation}
\label{higher1}
\Delta V(\phi) = -a_{j, k, l} M_*^{-j} \phi^{k} H_s^{l}.
\end{equation}
First we consider a case with $j, k \gg 1$ and $l=2$.
The potential is given by
\begin{eqnarray}
V(\phi) &=& V|_{\phi=M_*} + \int_{M_*}^{\phi} 
\left[\frac{d }{d\phi}\Delta V\right] d\phi\nonumber\\ 
&=& V|_{\phi=M_*}+\left[
 - a_{k-2, k, 2}M_*^{-(k-2)} \phi^{k} H_s^{2}
+a_{k-2, k, 2}M_*^{-(k-2)} M_*^{k} H_s^{2}
\right]\nonumber\\
&=& \hat{V}|_{\phi=M_*} - a_{k-2, k, 2}M_*^{-(k-2)} \phi^{k} H_s^{2},
\end{eqnarray}
where the definition of $\hat{V}$ is 
\begin{equation}
 \hat{V}\equiv V(\phi)+a_{j, k, l}M_*^{-j} \phi^{k} H_s^{l}.
\end{equation}
Assuming that the vacuum energy dominates $H_s$, we obtained
\begin{equation}
V(\phi)\simeq \frac{\hat{V}|_{\phi=M_*}}{1+a_{k-2, k, 2}A_v 
M_*^{-(k-2)} \phi^{k}M_p^{-2}}.
\end{equation}
At first, near $\phi \sim M_*$, this potential may appear to be a steep
function of $\phi$.
On the other hand, the approximate form of the potential in the region
 $\phi_1>\phi>\phi_c>M_*$ is given by 
\begin{equation}
V(\phi)\simeq \Lambda_* \phi^{-k},
\end{equation}
where $\phi_c$ and $\Lambda_*$ are defined as
\begin{equation}
\phi_c \equiv (M_*^{k-2}M_p^2/a_{k-2, k,2}A_v)^{1/k}
\end{equation}
and 
\begin{equation}
\Lambda_* \equiv \frac{\hat{V}|_{\phi=M_*}
 M_*^{k-2}M_p^{2}}{a_{k-2, k, 2}A_v},
\end{equation}
respectively.
Note that the above potential mimics quintessence.

Seeing the above results, one may think that our mechanism considered
here is the mechanism that explains the origin of the quintessential
potential, revealing especially the reason why quintessence is extremely
light and is rolling down to the vanishing energy state.
This speculation is true, but we have more to discuss about.

Before discussing ``more'' about our model,
let us further consider a case where $\Delta V$ in a region (for example)
$M_* <\phi_*<\phi$  is dominated by the correction with $l=4$, where
the potential is given by 
\begin{eqnarray}
V(\phi) &=& V|_{\phi=\phi_*} + \int_{\phi_*}^{\phi} 
\left[\frac{d }{d\phi}\Delta V\right] d\phi\nonumber\\  
&\simeq& \hat{V}|_{\phi=\phi_*} - a_{k, k, 4} M_*^{-k} \phi^{k}  
H_s^{4}.
\end{eqnarray}
Assuming that the vacuum energy dominates $H_s$, we obtained
\begin{equation}
V(\phi)\simeq \frac{P}{2}\left(
-1 + \sqrt{1+\frac{4 \hat{V}|_{\phi=\phi_*}}{P\phi^{-k}}}
\right)\phi^{-k},
\end{equation}
where $P$ is given by
\begin{equation}
P \equiv \frac{M_*^k M_p^4}{a_{k, k, 4}  A_v^2}.
\end{equation}
Here the solution with the negative sign is neglected because one cannot
reach at the negative solution.
Again, we obtained a potential that may explain the peculiar properties 
of quintessence.

\section{Coincidence problem?}
Let us explain another new idea, which may explain why the order of
the  
magnitude of the potential energy density is comparable to the energy
density in matter today. 
To be more precise, we will examine if one can obtain $\rho_v \sim
\rho_{m}$ in the above scenario, without using the ideas that have
been considered in quintessential cosmology.
What we will consider here is a case where the coefficient $A_m$
appears with the negative sign, which is the opposite to the sign of
$A_v$. 
For now, we will assume $|A_r| \ll 1$ and 
disregard the energy density in radiation $\rho_r$.\footnote{It is
straightforward to include $\rho_r$ in the above arguments.}
Then, the explicit form of $H_s^2$ is given by
\begin{equation}
H_s^2 = A_v \rho_v - A_m \rho_m,
\end{equation}
where $A_m$ is redefined to be positive.
In our scenario, the effective potential for the field $\phi$ is
destabilized if $H_s^2 > 0$ (i.e. $A_v \rho_v > A_m\rho_m$),
where the reduction of the potential energy proceeds through the
 mechanism that we have stated above.
On the other hand, if the energy density  in matter becomes large and
$H_s^2$ approaches to $H_s^2\simeq 0$, the reduction of the vacuum
energy is suspended.

If there is phase transition, the potential energy density may
(partially) be converted into the energy density in matter.
If the resultant energy density in matter is sufficiently large,
the field $\phi$ will ``rolls back'' the potential.

The reduction mechanism is active as far as the energy
density in matter satisfies the condition
\begin{equation}
\rho_v >\frac{A_m}{A_v}\rho_m,
\end{equation}
while the field $\phi$ becomes extremely light when
\begin{equation}
\rho_v \simeq\frac{A_m}{A_v}\rho_m,
\end{equation}
and then, it may start to dominate the energy density of the Universe 
in the later epoch.
The vacuum energy $\rho_v$ is positive and decreasing with time, 
and its later evolution is determined by the equation of motion
for the field $\phi$.

One may think that our mechanism considered here gives a ``reason'' to
the initial 
conditions that has been considered in quintessence cosmology.
This speculation is true, but we have another ``more'' to discuss
about.

Now, before discussing about another ``more'', 
let us consider energy density in radiation.
What happens if the Universe is dominated by radiation?
Assume that the coefficient $A_r$ is very small in a period of the
Universe, then the potential energy density at that time 
can be much smaller than the energy density in radiation
($\rho_r$) by a factor of $A_r/A_v$.
Then, while the energy density in radiation decreases with time,
the potential energy density decreases keeping the above factor.

It is easy to apply the above arguments to a realistic evolutionary
scenario of the Universe.
The evolution of the Universe after reheating will be
determined by the energy density in radiation, when the coefficient $A_r$
is very small ($|A_r|\ll 1$).
In a later epoch, when the energy density in matter becomes significant,
the potential energy density is bounded below by the ratio $A_m/A_v$.
More later evolution of the field $\phi$ is determined by the equation of
motion, which looks like quintessence.
In our model, the evolution of the field $\phi$ is not simple but very
sensitive to the changes in the Universe.
Of course, the coefficients may also vary with time, especially at phase
transitions when there are significant changes in the Universe.  
We will postpone the model-dependent analyses to our forthcoming paper.

\section{A dynamical approach}
There has been many proposals that might solve the cosmological constant
problem dynamically.
Here we consider a model named ``a dynamical approach'' that has been 
advocated by Mukohyama and Randall\cite{MR-Dyn}.
The novel feature is that a scalar field $\varphi$ exists which has
non-standard kinetic terms whose coefficient diverges at zero
curvature. 
The divergent coefficient of the kinetic term means that the lowest
energy state is never achieved, even if the minimum of the potential
seems to have negative energy.
Instead, the cosmological constant automatically stalls at or near
zero. 

To explain the basic idea of a dynamical approach, let us consider the
following action 
\begin{equation}
I= \int d^4x \sqrt{-g} \left[
\frac{R}{2\kappa^2} + \alpha R^2 + 
\beta \frac{K^q}{\kappa^4 f(R)^{2q-1}} - V(\varphi)
\right],
\end{equation}
where $K$ is a function given by $K=\kappa^4 \partial_{\mu} \varphi
\partial^{\mu} \varphi$.
Here it is assumed that $f(R)$ behaves as
\begin{equation}
f(R)\sim (\kappa ^2 R^2)^m.
\end{equation}
If $q>1/2$, the factor $\sim 1/f(R)^{2q-1}$ that appears in front of the
kinetic term becomes very large as $R$ approaches $0$.
As a result, the field $\varphi$ becomes extremely slow-roll and cannot
reach at the vacuum with vanishing energy density.

We found this interesting idea when we were considering about the relation
between our scenario and quintessence cosmology, and thought that there
must also be a relation between our scenario and a dynamical approach.
If so, we may find a unified picture of these
(seemingly different) approaches, which will provide us with a powerful
tool to settle the problems that arise in each model, through 
exchanging views among them.
In fact, in the above model of a dynamical approach, it is possible to
rescale the field $\varphi$ as  
$\varphi\rightarrow R^{m(2q-1)}\varphi$ to absorb the factor in front of
the kinetic term into the redefinition of $\varphi$, and then,
due to the redefinition of the field $\varphi$, the factor is moved into
the redefined potential $V(\varphi)$, which will be given by the form
 that is similar to eq.(\ref{higher1}).
Here we can use the relation between $R$ and $H$, which is given by
\begin{equation}
R= -6(2H^2 + \dot{H}),
\end{equation}
where $\dot{H}$ is negligible if $\varphi$ is a extremely slow-roll
field.
Obviously, the above relation and the redefinition of the field
$\varphi$ suggests that it is possible to reproduce our scenario from
the action that has been used in a dynamical approach.

Now it is very easy to find the solution to the coincidence problem, 
which was thought to be a serious problem in a dynamical approach. 
In fact, in a dynamical approach the problem was serious because
they have neglected other components 
(i.e, elements other than the scalar curvature $R$)
that may appear in $f$.
Of course, there is no reason to believe that other components
(for example, kinetic terms and potentials) that
appeared in the original action cannot appear in $f$.
Considering more generic forms of $f$, we will define a new function
$f(R_s)$, where a new variable $R_s$ 
is used instead of the scalar curvature $R$. 
$R_s$ is not a scalar curvature.
It is defined so that one can use the relation 
\begin{equation}
R_s \simeq  H_s^2.
\end{equation}
The new variable $R_s$ and the new function $f(R_s)$ modifies 
the scenario of a dynamical approach.
The most important discrepancy between the original scenario and our
modified one is that
the factor $\sim 1/f(R_s)^{2q-1}$ that appears in front of the
kinetic term becomes very large as $R_s$ (not $R$) approaches $0$.
As a result, the field $\varphi$ becomes extremely slow-roll and cannot
reach at the vacuum with $R_s=0$.
By analogy with our scenario, it is very clear that the coincidence
problem in a dynamical approach is not the problem in 
a dynamical approach itself, but the problem in the definition of
$f(R)$ and $V(\varphi)$.
To be precise, the original extension of the kinetic term
is not generic in a sense that it does not contain all the allowed
components in the theory.
Considering more generic extension of the kinetic term, one can 
obtain the solution to the coincidence problem in a dynamical approach,
using the mechanisms that we have discussed above in the previous
section.

\section{Conclusions and discussions}
Both inflationary and quintessence cosmologies require scalar fields
which roll slowly over cosmological time scales.
The condition seems quite severe in quintessence cosmologies, where
one needs to introduce an extremely slowly varying and light scalar field
that must roll toward the vanishing vacuum energy state.
In this paper, we started with a question whether it is possible to
``explain'' the above properties of quintessence cosmologies.

First, we showed from a new perspective that the conventional $O(H)$
corrections can be used to solve both the cosmological constant problem
and the coincidence problem. 
Besides the fact that the resultant potentials mimic quintessence,
we have a new idea to solve the coincidence problem, which is
qualitatively different from the usual techniques that has been used in 
quintessence cosmologies.

In this paper we considered $O(H)$ corrections to solve the problems,
while they are in some cases notorious for generating 
unwanted masses for the inflaton field.
For brane inflationary models, the inflationary scenario that 
has been discussed in ref.\cite{elliptic} may circumvent the
serious eta-problem.\footnote{In ref.\cite{elliptic} we mainly 
considered a model in which fluctuations are generated along the
equipotential surface, which is flat by definition. Application of
elliptic inflation to brane inflationary models is very interesting.
We also discussed in ref.\cite{elliptic} that $O(H)$ corrections lifting
the inflationary direction may not become a serious problem if
there is a symmetry enhancement at the tip of the inflationary throat.}
It is also notable that the mechanism presented in ref.\cite{elliptic} 
may induce a tilted spectrum that is determined by the inner
structure of the extra dimensions.
We are not going back to historical developments, however 
it is easy to understand that the idea of extra dimensions is very
attractive.
The idea of large extra dimensions is important, since it may solve or
weaken the hierarchy problem.
To be more precise, in models with large extra dimensions, fields in the
standard model(SM) are localized on a wall-like structure,  
while the graviton propagates in the bulk.
The discrepancy of the volume factor between gauge fields and
gravitational fields may explain the large hierarchy between gravity and
gauge interactions\cite{Extra_1}. 
An important progress in brane cosmology is made in
ref.\cite{brane-inflation0}, where a variant of D-term inflation is
found in the paradigm of brane cosmology\cite{angled-inflation,
matsuda_braneinflation}. 
In any case, it is an important challenge to find signatures of branes in 
cosmological observations.
Historically, this challenge has been made in a new paradigm of
brane defects\cite{brane-defects, matsuda-defects}.\footnote{Supergravity
provides us with a natural mechanism for removing cosmological domain
walls\cite{matsuda-wall}. }
Inflation models with low fundamental scale are important if extra
dimensions are large\cite{low_inflation, matsuda_thermalhyb}.
Obtaining the baryon asymmetry of the Universe in such low-scale models
is very difficult, but there are models that may solve this problem
\cite{low_baryo, Defect-baryo-largeextra,  
Defect-baryo-4D}.
In these models, cosmic defects play distinguishable roles and may leave
their signatures\cite{previous-onlystrings, matsuda_JGRG,
matsuda_necklace, BraneQball, matsuda_monopoles_and_walls, incidental, 
matsuda_angleddefect, tit-new}.
The curvatons\cite{curvaton_1, curvaton_2, curvaton_3} may also be
significant\cite{curvaton_liberate, topologicalcurvaton}, and is
expected to leave their signatures on CMB. 

In this paper, we considered $O(H)$ corrections that destabilize the
potential for the field $\phi$, and explained how one can obtain tiny
cosmological constant. 
Our model can explain why quintessence rolls slowly 
down to the vanishing energy state.
Another distinguishable property of our scenario is that we can explain
the 
reason why the order of the magnitude of the cosmological constant is
comparable to the energy density in matter today, without tuning the
initial conditions. 
Our result suggests that there will be no serious problem in obtaining
desired cosmological constant if the model has the properties that we
have discussed in this paper.
Unfortunately, it is still very hard to calculate the coefficients of
the higher corrections that are very important to examine the
model-dependent evolution.
We think further study on this issue is needed, which may reveal the 
particular reason why we are living in the Universe with extremely
tiny but substantial amount of cosmological constant.

Finally, we will make some comment about a principle that is required to
solve the cosmological constant problem.
The action that describes gravitational interactions seem to have
a shape that is recursively constructed by the elements that appears in
the usual formulation, which may be referred to as a fractal-like
formulation. 
For example, in a dynamical approach, the coefficient in front of a
kinetic term may depend on any element that appears in the original
action. 
As we have discussed in this paper, the idea is quite important 
because it may be used to reproduce the peculiar form of the
quintessential 
potential and also to explain why we are living with vanishingly small
but substantial amount of cosmological constant.
However, if one considers only the scalar curvature $R$, one will
find that the coincidence is a severe problem.
The coincidence problem may be solved if one includes ``other elements''
in the function, which is of course more generic than considering
only the scalar curvature $R$.
Any ``coefficient'' that appeared in the original action may be extended
to a generic function that depends on any element that appeared in the
original action.
As a result, the action is recursively constructed.
Extension of the original gravitational action has been discussed by
many authors.
As far as we know, this is the first attempt to explain
 successfully the cosmological constant problem and the coincidence
 problem, reproducing the famous quintessence cosmology and a dynamical
 approach from the simple setups.

\section{Acknowledgment}
We wish to thank K.Shima for encouragement, and our colleagues in
Tokyo University for their kind hospitality.

\end{document}